\documentstyle[twocolumn,prb,aps,epsfig]{revtex}
\draft
\begin{document}
\title{Superconductivity in Dense $MgB_2$ Wires}
\author{P. C. Canfield, D. K. Finnemore,  S. L. Bud'ko, J. E. Ostenson,  
G. Lapertot,\thanks{On 
leave from Commissariat a l'Energie Atomique, DRFMC-SPSMS, 
38054 Grenoble, France} C. E. Cunningham,\thanks{On leave from Dept. of Physics, 
Grinnell College, Grinnell, IA 50112} and C. Petrovic }
\address{Ames Laboratory, U.S. Department of Energy and Department of Physics and Astronomy\\
Iowa State University, Ames, Iowa 50011}
\date{\today}
\maketitle
\begin{abstract}
$MgB_2$ becomes superconducting just below 40 K. Whereas porous polycrystalline 
samples of $MgB_2$ can be synthesized from boron powders, in this letter we 
demonstrate that dense wires of $MgB_2$ can be prepared by exposing boron 
filaments to $Mg$ vapor.  The resulting wires have a diameter of 160 ${\mu}m$,  
are better than 80\% dense and manifest the full $\chi = -1/4{\pi}$ 
shielding in the superconducting state.  Temperature-dependent resistivity 
measurements indicate that $MgB_2$ is a highly conducting metal in the 
normal state with $\rho (40~K)$ = 0.38 $\mu Ohm$-$cm$.  Using this value, 
an electronic mean free path, $l \approx 600~\AA$ can be estimated, 
indicating that $MgB_2$ wires are well within the clean limit. $T_c$, $H_{c2}(T)$, 
and $J_c$ data indicate that $MgB_2$ 
manifests comparable or better superconducting properties in dense wire 
form than it manifests as a sintered pellet.    
\end{abstract}

\section{Introduction}

The discovery of superconductivity in $MgB_2$ has caused a renaissance of 
interest in intermetallic superconductivity.\cite {1} This, combined with the 
discovery of superconductivity in $YPd_2B_2C$ and the $RNi_2B_2C$ series several 
years ago,\cite {2-5} seems to indicate that the old idea of 
looking for high intermetallic $T_c$ values in compounds rich in light elements is still a 
valid guiding principle. Measurements of the boron isotope effect in this 
compound\cite {6} are consistent with the superconductivity being mediated via 
electron-phonon coupling, a conclusion that is also suported by 
recent bandstructural calculations.\cite {7} Measurements of the upper 
critical field, $H_{c2}(T)$, the thermodynamic critical field, $H_c(T)$, and the 
critical current, $J_c$, indicate that $MgB_2$ is a type-II superconductor with 
properties that are consistent with a intermetallic superconductor that has a 
$T_c \approx 40~K$.\cite {8} For example, other than the 
remarkably high $T_c$, $MgB_2$ appears to be 
quite similar to $Nb_3Sn$. Given this similarity, and given the 
far lower density of $MgB_2$ as well as the greater natural abundances of $Mg$ and $B$, 
a logical question is whether wires of $MgB_2$ can be easily synthesized, and 
if so what are their physical properties.  In this letter we present a 
remarkably simple method for the synthesis of $MgB_2$ wires from boron 
filaments. In addition, we show that wires produced in this manner are 
of high density and have impressively low normal state resistivity 
and impurity scattering.

\section{Experimental Methods}

$MgB_2$ can be synthesized in powder form by reacting stoichiometric amounts 
of powdered $B$ and $Mg$ at 950$^{\circ}$C for approximately an hour.\cite {6} 
Given that at 950$^{\circ}$C the vapor pressure of $Mg$ is 
approximately 200 Torr,\cite {9} it is believed that $MgB_2$ forms 
via a process of diffusion of $Mg$ vapor into the boron grains. 
Based on this observation, the possibility of using this technique 
on other morphologies of boron appears to be promising. 

$MgB_2$ wire was produced by sealing 100 ${\mu}m$ diameter boron 
fiber\cite {10} and $Mg$ into a $Ta$ tube with a nominal ratio of $Mg_2B$. 
Given that $MgB_2$ is the most $Mg$ rich 
binary $Mg-B$ compound known,\cite {11} it was felt that excess $Mg$ would aid in the 
formation of the proper, stoichiometric phase. The sealed $Ta$ tube was sealed 
in quartz and then placed into a 950$^{\circ}$C box furnace for 
approximately an hour. The reaction ampoule was then removed from 
the furnace and quenched to room temperature. 

Measurement of temperature and field dependent electrical resistivity and 
magnetization were performed in Quantum Design MPMS and PPMS systems. 
Resistivity measurements were made in a standard four probe geometry using 
Epotek H20E silver epoxy to make contacts. The contact resistance was 
approximately 1 Ohm.  Given the well defined geometry of the samples, accurate 
measurements of resistivity were possible.

\section{Results}
Upon opening the $Ta$ tube it became clear that there had been a reaction 
between the boron fiber and the $Mg$ vapor. Whereas the boron fibers were straight 
and moderately flexible before the reaction, the $MgB_2$ wires in the $Ta$ tube after 
the reaction were brittle and deformed. The inset of Figure 1 is a photograph of 
the resulting wires. As can be seen, there has been significant warping 
and bending of the fiber as a result 
of the reaction with the $Mg$ vapor. Figure 2 shows scanning electron 
microscope images of the fiber before the reaction as well as the wire after 
the reaction. In both cases a tungsten core (approximately 15 ${\mu}m$ 
diameter) can be clearly seen. This core is part of the original boron fiber 
and does not appear to be effected by the exposure of the fiber to $Mg$, nor, 
as will be seen, does it seem to effect the superconducting properties of the 
resulting $MgB_2$ wire. Whereas the boron fiber has a diameter of 100 ${\mu}m$ and 
breaks with a smooth, clean surface (inset to Fig. 1), the $MgB_2$ wire has a diameter of 
approximately 160 ${\mu}m$ and breaks with a rougher, grainier surface. The 
increased diameter of the wire is consistent with observations that there is 
an expansion associated with the formation of the $MgB_2$ powders during 
synthesis.\cite {6} Although the $MgB_2$ wires are somewhat brittle, the integrity 
of the filament segments was preserved during the exposure to the $Mg$ vapor, 
i.e. the fibers did not decompose or turn into powder.

Using a diameter of 160 ${\mu}m$ and measuring the length and mass 
of several wire segments we determined the density of the 
wire to be 2.4 $g/cm^3$. This is to be 
compared with a theoretical value of 2.55 $g/cm^3$ 
for a single crystal sample using 
lattice parameters $a$ = 3.14 $\AA$ and $c$ = 3.52 $\AA$ for the hexagonal 
unit cell.\cite {6} Given 
the rather coarse nature of our measurements, this implies that the wire 
samples are probably better than 80\% of the theoretical density.  It should 
be noted that the small tungsten core would come in as a roughly 10\% 
correction, and therefore is within our level of uncertainty.

Figure 1 presents the temperature-dependent magnetization of $MgB_2$. The 
data were taken after cooling in zero field and then warming in a field of 25 
Oe. Given the aspect ratio of the wire segments used we were able to obtain 
a susceptibility very close to $-1/4{\pi}$, the value expected for total shielding 
and a demagnetization factor close to zero. $T_c$ = 39.4 K can be determined from these 
data by using an onset criterion (2\% of $-1/4{\pi}$). The width of the transition 
(10\% - 90\%) is 0.9 K. 

Figure 3 presents the temperature-dependent electrical resistivity of 
$MgB_2$ wires. The room temperature resistivity has a value of 9.6 ${\mu}Ohm$-$cm$ whereas 
$\rho (77~K)$ = 0.6 ${\mu}Ohm$-$cm$ and $\rho (40~K)$ = 0.38 ${\mu}Ohm$-$cm$. 
This leads to a residual resistivity ratio 
of $RRR$ = 25.3. It should be noted that the shape of the resistivity curve and the 
$RRR$ values are qualitatively the same as those observed for sintered pellets 
of polycrystalline $Mg^{10}B_2$.\cite {8} The relatively low room temperature 
resistivity value, along with the high $RRR$ are not unusual for diboride 
samples.\cite {12} The resistivity of the sintered pellet samples\cite {8}
is approximately 1 ${\mu}Ohm$-$cm$ at 40 K. This somewhat higher 
value of the calculated resistivity for the pellet is consistent with the sintered 
sample having an actual density substantially lower than the theoretical value.

The temperature-dependent resistivity shown in Fig. 3 can be fit by $\rho = 
 \rho_0 + \rho_1 T^{\alpha}$ with $\alpha \approx 2.6$ between $T_c$ and 
200 K.  This is comparable to the power law $R~=~R_1 + R_1 T^{\alpha_1}$ 
with $\alpha_1 \approx 2.8$
found for the sintered $Mg^{10}B_2$ sample\cite {8} over 
a comparable temperature range.
Given the similarity of the two power laws it seems clear that the 
resistivity of $MgB_2$ will not have a linear slope for temperatures between 
$T_c$ and 300 K.\cite {7}  On the other hand, using an average Fermi 
velocity\cite {7} of $v_F = 4.8 \cdot 10^7~cm/s$ and a carrier density of $6.7 \cdot 10^{22} 
~el/cm^3$ (two free electrons per unit cell) 
we can estimate the electronic mean free path to be approximately 
600 $\AA$ at $T_c$.  This is clearly an approximate value of the electronic mean 
free path, but given the estimated superconducting coherence length of 
approximately 50 $\AA$,\cite {8} these values place 
$MgB_2$ wires well within the 
clean limit.  Given a $\kappa \approx 26$,\cite {8} this implies that, 
much like the case of the $RNi_2B_2C$ 
materials,\cite {5} there may be significant non-local effects 
associated with $MgB_2$.

The superconducting transition temperature, $T_c$ = 39.4 K, can be determined from 
both the magnetization and resistivity data shown in Figs. 1 and 3. This value 
is slightly higher than the $T_c$ = 39.2 K value determined for isotopically 
pure $Mg^{11}B_2$, but is significantly lower that $T_c$ = 40.2 K for $Mg^{10}B_2$. 
This is consistent with an approximate 80 \% natural abundance of $^{11}B$. 
It is noteworthy that the superconducting transition is both relatively high 
and sharp in the wire samples. This means that either very few impurities are 
being incorporated into the $MgB_2$ or that what few impurities are being 
incorporated are having very little effect on either resistivity or $T_c$.

The temperature dependence of the upper critical field, $H_{c2}(T)$, is shown 
in the inset to Fig. 3. For each field three data points are shown: onset 
temperature, temperature for maximum $d{\rho}/dT$, and completion temperature. 
Qualitatively these data are remarkably similar to the $H_{c2}(T)$ data 
inferred from measurements on $Mg^{10}B_2$ sintered pellets\cite {8} 
as well as recent measurements on hot-pressed powders.\cite {13} 
Quantitatively, 
at $H$ = 9 T, the width of the resistive transition for the wire sample 
is roughly half of the width found for the sintered sample. 
These data are consistent 
with the wire sample being of comparable or better quality as the sintered 
powder samples. 

Figure 4 presents data on the critical current $J_c$. The open symbols are $J_c$
values extracted from direct measurements of the current dependent voltage 
across the sample at given temperature and applied field values. The 
filled symbols are $J_c$ values inferred from magnetization loops by 
application of the Bean model.\cite {8,14} The direct measurement 
of $J_c$ was limited to values below approximately 200 $A/cm^2$ due to resistive 
heating from the sample leads and contact resistance. As can be seen, the 
extrapolations of the directly measured, low $J_c$, data and the 
Bean-model-inferred, high $J_c$, data match up moderately well. 
In comparison to the $J_c$ data presented for a sintered pellet of 
$Mg^{10}B_2$,\cite {8} $J_c$ for the wire sample is roughly a factor 
of two higher at low fields and temperatures and over an order of magnitude 
higher at high fields.

\section{Conclusions}

We have devised a simple technique of producing low resistivity, high density, 
high $T_c$ $MgB_2$ in wire form via exposure of boron filaments to $M$g vapor. 
The resulting wire has better than 80\% the theoretical density of 
$MgB_2$ and measurements 
of the temperature dependent resistivity reveal that $MgB_2$ is highly conducting 
in the normal state. The room temperature resistivity has a value of 
9.6 ${\mu}Ohm$-$cm$ whereas the resistivity at $T$ = 40 K is 0.38 ${\mu}Ohm$-$cm$.
This means that even in the normal state wires of $MgB_2$ can carry 
significant current densities. This should be compared with the resistivity of $Nb_3Sn$ 
$\rho (20~K)$ = 11 $\mu Ohm$-$cm$ and $\rho (300~K)$ = 80 $\mu Ohm$-$cm$.\cite {15}

Given the well-defined geometry of the wire samples we have been able 
to directly measure a full $-1/4{\pi}$ susceptibility in the superconducting 
state. The values of $T_c$ for this wire sample are slightly higher than 
the $T_c$ values for isotopically pure $Mg^{11}B_2$ sintered powders and the 
width of the resistive superconducting transition is smaller that that 
seen for $Mg^{10}B_2$ sintered powders. In addition, $H_{c2}(T)$ for the 
wire sample is virtually the same as that found for $Mg^{10}B_2$ sintered 
pellets. Based on all of these observations it appears that $MgB_2$ wires 
provide dense, high quality samples of $MgB_2$. By comparing our estimate 
of the electronic mean free path $l \approx 600~\AA$ to the 
superconducting coherence length $\xi \approx 50~\AA$\cite {8} 
we can see that $MgB_2$ wires are well within the clean limit.

All of the above, of course, presents the possibilities of using 
such wires for both research 
and applied purposes. For basic research the possibilities of making weak link 
Josephson junctions, SLUGS, and other devices are currently being pursued. 
On the applied side, given that boron filaments are produced in a variety of sizes 
and of arbitrary lengths, the possibility of converting boron filament into 
$MgB_2$ wire as part of a continuous process leads to the possibility of 
simple manufacturing of light weight, high $T_c$, wires with remarkably small 
normal state resistivities. In addition, this process could be used to 
turn boron coatings on tapes, cavities, or other devices into 
high-quality superconducting films. Although the low 
temperature $J_c$ values are currently smaller than those for 
$Nb_3Sn$,\cite {8} as of yet very little effort has been put 
into optimizing $J_c$. A multi-dimensional phase space of filament 
purity, diameter, treatment time and temperature has yet to be explored. 
Both basic and applied directions of research will have to be explored in 
detail over the coming months and years.

\section{Acknowledgments}

We would like to thank V. Crist and M. Kramer for taking the electron 
microscope images as well as J. R. Clem and R. W. McCallum for 
useful discussions and N. Anderson 
and N. Kelso for assistance in the lab. Ames Laboratory is operated for 
the U. S. Department of Energy by Iowa State University under Contract 
No. W-7405-Eng.-82. This work was supported by the director fo Energy 
Research, Office of Basic Energy Sciences.

\begin{figure}
\epsfxsize=0.9\hsize
\vbox{
\centerline{
\epsffile{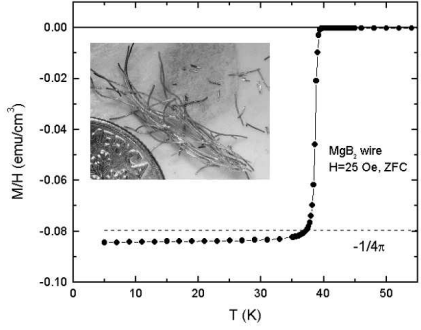}
}}
\caption{Magnetization divided by applied field (25 $Oe$) for zero field 
cooled wire sample.  Field was applied parallel to the wire length, leading 
to a demagnetization factor close to zero.  Inset:  photograph of wires as 
they appear after removal from $Ta$ tube and part of U. S. dime for scale.} 
\label{F1}
\end{figure}
\begin{figure}
\epsfxsize=0.9\hsize
\vbox{
\centerline{
\epsffile{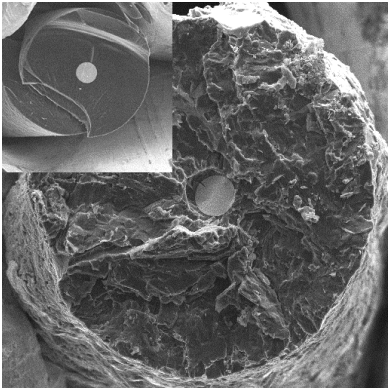}
}}
\caption{Electron microscope image of cross section of grown $MgB_2$ 
wire.  The diameter of the wire is 160 ${\mu}m$.  Inset:  electron microscope 
image of the un-reacted boron filament.  The diameter of the filament is 
100 ${\mu}m$.  For both images the wire / filament was snapped 
in-situ.  Note:  In both images a central core of tungsten wire 
(diameter $\approx 15~{\mu}m$) can 
be clearly seen. }
\label{F2}
\end{figure}
\begin{figure}
\epsfxsize=0.9\hsize
\vbox{
\centerline{
\epsffile{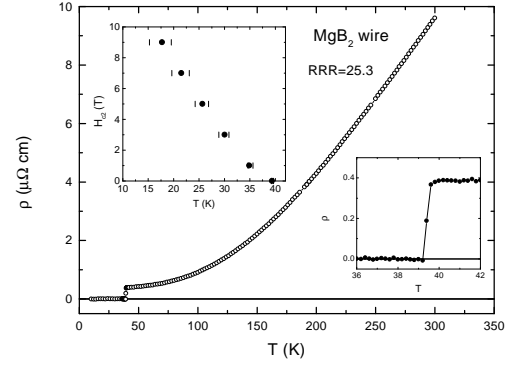}
}}
\caption{Temperature dependent electrical resistivty of $MgB_2$ wire.  Lower 
inset:  Expanded view for temperatures near $T_c$.  Upper inset:  $H_{c2}(T)$ data 
inferred from temperature dependent resistivity data taken in constant 
applied field upon cooling.  The three symbols are for onset, 
maximum slope and completion temperatures. }
\label{F3}
\end{figure}
\begin{figure} 
\epsfxsize=0.9\hsize
\vbox{
\centerline{
\epsffile{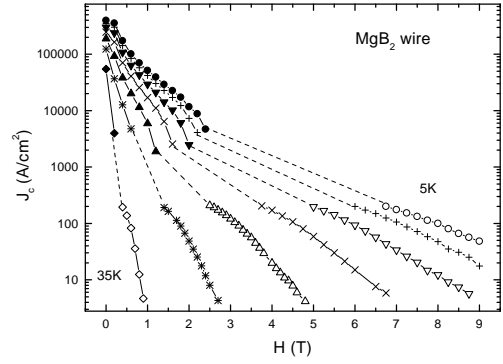}
}}
\caption{Superconducting critical current density, $J_c$, as a function of 
applied field every 5 K in 5 - 35 K range.  Open symbols were 
taken via direct measurement of current dependent voltage of the wire.  Filled 
symbols were determined via a Bean model analysis of 
magnetization data from wire samples with the applied field parallel to the 
wire length.  The dashed lines simply connect data sets taken at the same 
temperature.}
\label{F4}
\end{figure}

\begin{references}  
\bibitem{1} J. Akimiitsu, Symposium on Transition Metal Oxides, Sendai, January 10, 2001;
J. Nagamatsu, N. Nakagawa, T. Muranaka, Y. Zenitani, and J. Akimitsu (to be published).
\bibitem{2} R. Nagarajan, C. Mazumdar, Z. Hossain, S. K. Dhar, 
K. V. Gopalakrishnan, L. C. Gupta, C. Godart, B. D. Padalia, and 
R. Vijayaraghavan, Phys. Rev. Lett. {\bf 72}, 274 
(1994).
\bibitem{3} R. J. Cava, H. Takagi, B. Battlog, H. W. Zandbergen, J. J. Krajewski,
W. F. Peck, Jr., R. B. van Dover, R. J. Felder, T. Siegrist,  
K. Mizuhashi, J. O. Lee, H. Eisaki, S. A. Carter, and S. Uchida, 
Nature {\bf 376}, 146 (1994).
\bibitem{4} R. J. Cava, H. Takagi, H. W. Zandbergen, J. J. Krajewski,
W. F. Peck, Jr., T. Siegrist, B. Battlog, R. B. van Dover, R. J. Felder, 
K. Mizuhashi, J. O. Lee, H. Eisaki, and S. Uchida, 
Nature {\bf 376}, 252 (1994).
\bibitem{5} For review see: P. C. Canfield, P. L. Gammel, and D. J. Bishop, 
Physics Today {\bf 51}, 40 (1998) and references therein.
\bibitem{6} S. L. Bud'ko, G. Lapertot, C. Petrovic, C. E. Cunningham, N. Anderson, 
and P. C. Canfield, Phys. Rev. Lett. (in press); cond-mat/0101463.
\bibitem{7} J. Kortus, I. I. Mazin, K. D. Belashchenko, V. P. Antropov, and L. L. Boyer, 
cond-mat/0101446. 
\bibitem{8} D. K. Finnemore, J. E. Ostenson, S. L. Bud'ko, 
G. Lapertot, and P. C. Canfield, cond-mat/0102114.
\bibitem{9} R. E. Honig, and D. A. Kramer, Vapor Pressure Curves 
of the Elements, Fall 1968, RCA Laboratories, Princeton N.J. 
\bibitem{10} Textron Systems, 201 Lowell St., Wilmington, MA 01887. 
\bibitem{11} Binary Alloy Phase Diagrams, Second Edition, 
Edited by T. Massalski, (A.S.M. International, 1990).
\bibitem{12} A. N. Emel'yanov, V. P. Kobyakov, V. I. Ponomarev, 
T. G. Utkina, and A. S. Shteinberg, Phys. Met. Met. {\bf 83}, 608 (1997).
\bibitem{13} D. C. Larbalestier, M. Rikel, L. D. Cooley, A. A. Polyanskii, J. Y. Jiang, 
S. Patnaik, X. Y. Cai, D. M. Feldmann, A. Gurevich, A. A. Squitier, M. T. 
Naus, C. B. Eom, E. E. Hellstrom, R. J. Cava, K. A. Regan, N. Rogado, M. A. 
Hayward, T. He, J. S. Slusky, P. Khalifah, K. Inumaru, and M. Hass, 
unpublished.
\bibitem{14} C. P. Bean, Phys. Rev. Lett. {\bf 8}, 225 (1962).
\bibitem{15} Study of Transition Temperatures in Superconductors, 
Final Report, prepared by L. J. Vieland, and R. W. Cohen, RCA 
Laboratories, Princeton, NJ 08540, 1970.
\end{references}
\end{document}